\documentclass[prl,notitlepage,superscriptaddress,showpacs]{revtex4-1}
\usepackage{amsmath,amsfonts,amsbsy,amssymb,amsthm,bm,bbold}
\usepackage[pdftex]{graphicx}
\usepackage{epstopdf}

\newcommand{\comment}[1]{}
\newcommand{\tr}{{\rm Tr}}
\newcommand{\rk}{{\rm rank}}

\newcommand{\e}{{\rm e}}
\newcommand{\nn}{\nonumber\\}
\newcommand{\ket}[1]{| #1 \rangle}
\newcommand{\bra}[1]{\langle #1 |}

\newcommand{\up}{\uparrow}
\newcommand{\down}{\downarrow}

\begin{document}

\title{Quantum Tomography Protocols with Positivity are Compressed Sensing Protocols}
\author{Amir Kalev}
\affiliation{Center for Quantum Information and Control, University of New Mexico, Albuquerque, NM 87131-0001, USA}
\author{Robert L. Kosut}
\affiliation{SC Solutions, 1261 Oakmead Parkway, Sunnyvale, California 94085, USA}
\author{Ivan H. Deutsch}
\affiliation{Center for Quantum Information and Control, University of New Mexico, Albuquerque, NM 87131-0001, USA}

\begin{abstract}
Characterizing complex quantum systems is a vital task in quantum information science.  Quantum tomography,  the standard tool used for this purpose, uses a well-designed measurement record to reconstruct quantum states and processes.  It is, however, notoriously inefficient.  Recently, the classical signal reconstruction technique known as  ``compressed sensing" has been ported to quantum information science to  overcome this challenge: accurate tomography can be achieved with substantially fewer measurement settings, thereby greatly enhancing the efficiency of quantum tomography.  Here we show that compressed sensing tomography of quantum systems is essentially guaranteed by a special property of quantum mechanics itself---that the mathematical objects that describe the system in quantum mechanics are matrices with nonnegative eigenvalues.  This result has an impact on the way quantum tomography is understood and implemented.  In particular, it implies that the information obtained about a quantum system through compressed sensing methods exhibits a new sense of  ``informational completeness.''  This has important consequences on the efficiency of data taking for quantum tomography, and enables us to construct informationally complete measurements that are robust to noise and modeling errors.  Moreover, our result shows that one can expand the numerical tool-box used in quantum tomography and employ highly efficient  algorithms developed to handle large dimensional matrices on a large dimensional Hilbert space. While we mainly present our results in the context of quantum tomography, they apply to the general case of positive semidefinite matrix recovery.
\end{abstract}

\maketitle

\section{introduction}
Determining an unknown signal from a set of measurements is a fundamental problem in 
science and engineering.  However, as the number of free parameters defining the signal 
increases, its tomographic determination may become a daunting task.  Fortunately, in many contexts there is prior information about the signal that may be useful for tomography.  
Compressed sensing~\cite{donoho06, candes06a, candes06b, candes08, candes09, 
candes10a, candes10b, recht10, candes11} is a signal recovery technique developed for this aim.  It utilizes specific types of prior information about the structure of the signal to 
substantially compress the amount of information needed to reconstruct it with high 
accuracy.  In particular, it harnesses the prior information that the signal has a concise 
representation, e.g., that it is a sparse vector with a few nonzero elements or a low-rank 
matrix with a few nonzero singular values.  The  compressed sensing protocol then defines special classes of  measurements, henceforth referred to as ``compressed sensing measurements,'' that enable the unique identification of the signal from within the restricted set of  sparse vectors or low-rank matrices using substantially fewer measurement settings.  Moreover, it provides algorithms for efficient reconstruction by defining a specific class of convex optimization heuristics  whose solution determine the unknown signal from the measurement outcomes with very high accuracy (see Methods).  Importantly, solving any other optimization programs outside this class will not necessarily result in a compressed sensing protocol.

In the context of quantum information science, the ``signals" we seek to reconstruct are, 
for example, quantum states and processes, and the protocol for reconstruction is  quantum tomography.  
Because the number of free parameters in quantum states and 
processes scale poorly (growing as some power of the total Hilbert space dimension, 
which in turn grows exponentially with the number of subsystems), there has been a 
concerted effort to develop techniques that minimize the resources necessary for 
tomography.  To this end, the methodology of compressed sensing has been applied to 
the problem of quantum tomography~\cite{kosut08, gross10,liu11, shabani11, flammia12, 
liu12, sanders12, smith13, schwemmer14, rodionov14, tonolini14,kueng14}.  

In the pioneering work of ~\cite{gross10, liu11,shabani11} it was proved that 
quantum measurements can be easily designed to be within the special class of measurements 
required for compressed sensing.  Then, using the specifically chosen convex optimization, low-rank density matrices (close to pure quantum states) or low-rank process matrices (close to unitary evolutions) can be accurately reconstructed with a substantially reduced 
number of measurement settings.

The work we report here identifies a critical link between quantum tomography and compressed sensing. We discuss in particular the case of quantum state tomography, where the aim is to recover the density matrix, a {\em positive semidefinite} matrix, typically normalized with unit trace.  We show that the positivity property alone imposes a powerful constraint that places strong restrictions on the physical states that are consistent with the data.   As illustrated in Fig.~\ref{fig:sets}, this restriction is stronger than the one present in generic compressed sensing of  signals which are not necessarily positive semidefinite matrices.  This, in turn, has far reaching consequences.  First and foremost, it implies that as long as quantum measurements are within the special class associated with compressed sensing, then any optimization heuristic that contains the positivity constraint is effectively a compressed sensing protocol.  Second, tools provided by the compressed sensing  methodology now enable the construction of special types of informationally complete measurements that are robust to noise and to small model imperfections, with rigorous bounds. Finally, our results fundamentally unify many different quantum tomography protocols which were previously thought to be distinct, such as maximum-likelihood solvers, under the compressed sensing umbrella. We emphasize that constraining the normalization (trace) to a fixed value, as one does for density matrices, plays no role in the theorems we discuss below. Thus our results extend beyond the context of quantum state tomography, applying, e.g., to process tomography when the latter is described by a completely positive map, and more generally  to the reconstruction of low-rank positive semidefinite matrices.

\section{results}

\begin{center}\bf{Informational completeness}\end{center}

In quantum theory, a measurement is represented by a positive operator-valued measure, POVM, a set of positive semidefinite $d\times{d}$ matrices that form a resolution of the identity,  ${\mathcal E}=\{E_\mu|E_\mu\geq0,\sum_\mu E_\mu=\mathbb{1}\}$. The elements of a POVM represent the possible outcomes (events) of the measurement, and probability of measuring an outcome $\mu$ is given by the usual Born rule, $p_\mu = \tr(E_\mu \rho)$, where $\rho$ is the state of the system, a positive semidefinite matrix, $\rho\geq0$, normalized such that $\tr\rho=1$.  In the context of quantum-state tomography,  {\em informationally complete}  measurements play a central role. Let $\mathcal{S}$ be the set of all quantum states (density matrices). A measurement is said to be informationally complete if~\cite{scott06}
\begin{equation} 
\label{IC_definition}
\forall \rho_{a}, \rho_b \in \mathcal{S}, \;\rho_{a}\neq \rho_b, \, \, \exists E_\mu\in{\mathcal E} \,\, {\rm {\rm s.t.}} \,\, \tr(E_\mu \rho_a) \neq \tr(E_\mu \rho_b).
\end{equation}
In other words, no two distinct states $\rho_a$ and $\rho_b$ yield the same measurement outcome probabilities. Thus, a (noise-free) record of an informationally complete measurement uniquely determines the state of the system.  In general, for a $d$-dimensional Hilbert space, an informationally complete measurement consists of at least $d^2$ outcomes (POVM elements). 

While Eq.~\eqref{IC_definition} gives a general definition of an informationally complete measurement, if one has prior information about the state of the system, we can make this definition more specific~\cite{heinosaari13,carmeli14}. In particular, suppose the state is known {\em a priori} to be of a special class, ${\cal P}$, e.g., the class of density matrices of at most rank $r$. One defines a measurement to be ${\cal P}$ restricted informationally complete (restricted-IC)  if it can only uniquely identify a quantum state from within the subset ${\cal P}$, but cannot necessarily uniquely identify it from within the set of all quantum states. 
Such  ${\cal P}$ restricted-IC measurements can be composed of fewer outcomes than the $d^2$ outcomes required for a general informationally complete measurement. For example, Heinosaari~{\em et al.}~\cite{heinosaari13} showed that when ${\cal P}$ is the set of density matrices of at most rank $r$, then rank-$r$ restricted-IC measurements can be constructed with ${\cal O}(rd)$ outcomes, rather than ${\cal O}(d^2)$ outcomes required for a general informationally complete measurement.  One can formalize this definition in the context of quantum-state tomography. A measurement is said to be ${\cal P}$ restricted-IC, if~\cite{heinosaari13}
\begin{equation} 
\label{restricted_IC_definition}
\forall \rho_{a},\rho_b\in{\cal P}, \rho_{a}\neq \rho_b, \, \, \exists E_\mu\in{\mathcal E} \,\, {\rm {\rm s.t.}} \,\, \tr(E_\mu \rho_a) \neq \tr(E_\mu \rho_b).
\end{equation}
In some situations, a measurement can satisfy a stricter definition of informational completeness than the ${\cal P}$ restricted-IC of Eq.~\eqref{restricted_IC_definition}. A measurement is said to be ${\cal P}$ strictly-IC, if~\cite{carmeli14}
\begin{equation} 
\label{strictly_IC_definition}
\forall \rho_{a}\in{\cal P}, \forall\rho_b \in \mathcal{S},\; \rho_{a}\neq \rho_b, \, \, \exists E_\mu\in{\mathcal E} \,\, {\rm {\rm s.t.}} \,\, \tr(E_\mu \rho_a) \neq \tr(E_\mu \rho_b).
\end{equation}
There is a subtle yet important difference in the definitions of  ${\cal P}$ restricted-IC and   ${\cal P}$ strictly-IC. While the measurement record of the former identifies a unique state within the set ${\cal P}$, the measurement record of a the latter identifies a unique state within the set of {\em all} quantum states.  These notions of informationally completeness are key to understanding compressed sensing and its application in quantum tomography, as we discuss below.

\begin{center}{\bf The relation between informational completeness and compressed sensing}\end{center}
At its heart, the compressed sensing methodology employs prior information to reduce the number of measurements required to reconstruct an unknown signal.  Here we consider the compressed sensing recovery of a $d\times d$ Hermitian matrix, $M$.  Let the measurement record be specified as a vector-valued linear map, $y_i = {\cal A}[M]_i = \tr(A_i M)$, where ${\cal A}$ is known as the ``sensing map."  In general, when the set $\{A_i \}$ forms a basis for $d\times d$ matrices with at least $d^2$ elements~\cite{dariano04}, then the measurement record is informationally complete in the sense of Eq.~\eqref{IC_definition}, and in the absence of measurement noise, the signal can be recovered uniquely.  

If, however, we know {\em a priori} that $\rk(M)\leq{r}$, with $r\ll  d$, then we can substantially reduce the number of measurement samples required to uniquely reconstruct the unknown signal-matrix.  This is codified in a theorem by Recht {\em et al.}~\cite{recht10} and Cand{\`e}s {\em et al.}~\cite{candes11} which we restate as follows:\vspace{5pt}\\
{\bf Theorem~[compressed sensing].}  
Let the unknown signal $M_0$ be a Hermitian matrix with $\rk(M_0)\leq r$, and let $\bm{y}={\cal A}[{M_0}]$ be the measurement record obtained by a sensing map, ${\cal A}$, that  corresponds to compressing sensing measurements  for rank $r$.  Then $M_0$ is the {\em unique} Hermitian matrix within the set of {\em low-rank} Hermitian matrices (up to rank $r$) that is consistent with the measurement record. \vspace{5pt}\\
Importantly, in compressed sensing, when $r\ll d$, there are generally an infinite number of Hermitian matrices with rank larger than $r$ that are consistent with the measurement record. Thus,  the measurement record  associated with compressed sensing cannot uniquely specify $M_0$ among all $d\times d$ Hermitian matrices, and therefore it is not informationally complete in the sense of Eq.~\eqref{IC_definition}.  If, however,  the sensing map ${\cal A}$  corresponds to compressed sensing measurements (e.g., it satisfies the restricted isometry property~\cite{candes08}, see Methods), then according to the above theorem, the measurement record uniquely specifies $M_0$  within the restricted set of low-rank Hermitian matrices (rank$(M)\leq r\ll d$). Therefore compressed sensing measurements correspond to rank-$r$ restricted-IC, in the sense of Eq.~\eqref{restricted_IC_definition}.

This relation between compressed sensing measurements and rank-$r$  restricted-IC implies that any successful search must be restricted to  the low rank set of Hermitian matrices.  To achieve this, one solves the convex optimization problem~\cite{recht10,candes11,kueng14},
\begin{equation}
\label{CS_matrix}
\hat{M} = \arg\min_{M} \Vert{M}\Vert_*\,\, \, {\rm s.t.}\, \, \bm{y}={\cal A}[M],
\end{equation}
where $\Vert{M}\Vert_*=\tr\sqrt{M^\dag M}$, is the nuclear (or trace) norm, which serves as the convex proxy for rank minimization. Under the conditions above, the optimal solution is $\hat{M}=M_0$, i.e., exact recovery.  The use of the nuclear norm is essential here. If one uses only the compressed number of samples, solving any other optimization  that is not related to the above rank-minimization heuristic by some regularization will not result in a successful recovery. For example, the solution of the convex  programs $\arg\min_M \tr(M) \,\, \, {\rm s.t.}\, \, \bm{y}={\cal A}[M]$, and  $\arg\min_M \Vert \bm{y}-{\cal A}[M]\Vert_2$ with $m \ll {d^2}$ samples $\{y_i\}$ will generally yield a solution that is very different from $M_0$.  Such estimators generally require $m\sim{d^2}$ samples to recover $M_0$. The analogous result holds for compressed sensing of sparse vectors.  There ones require minimization of the $\ell_1$ norm of the vector, a convex heuristic for vector-sparsity.

In what follows, we specialize the compressed sensing paradigm to the case of  positive matrix recovery, and particular to quantum-state tomography.  There, the aim is to recover the state of the system, $\rho$, which has the key property of  positivity, $\rho\geq0$. 

\begin{center}{\bf The role of positivity in compressed sensing quantum tomography}\end{center}

Our central result is summarized in the following theorem:\vspace{5pt}\\
{\bf Theorem~1.} Let  $P_0$ be a positive semidefinite matrix with $\rk(P_0)\leq r$, and let $\bm{y}={\cal A}[{P_0}]$ be the measurement record obtained by a sensing map ${\cal A}$ that corresponds to compressing measurements  for a rank-$r$ Hermitian matrix.  Then $P_0$ is the {\em unique} matrix within the set of  positive semidefinite matrices of {\em any rank} that is consistent with the measurement record.\vspace{5pt}\\
This is an analogous theorem to the one presented by Bruckstein {\em et al.}~\cite{bruckstein08} for the case of positive sparse vector solutions for an underdetermined set of linear equations. Its proof as well as the details concerning the requirements on the   sensing map are given in the Supplementary information Section~A. It also extends a result by Cand{\`e}s {\em et al.}~\cite{candes13} and Demanet and Hand~\cite{demanet14} from rank-1 matrices to matrices with rank $\leq r$ for all permissible $r$.   

\begin{figure}[t]
\centering
\includegraphics[width=0.6\linewidth]{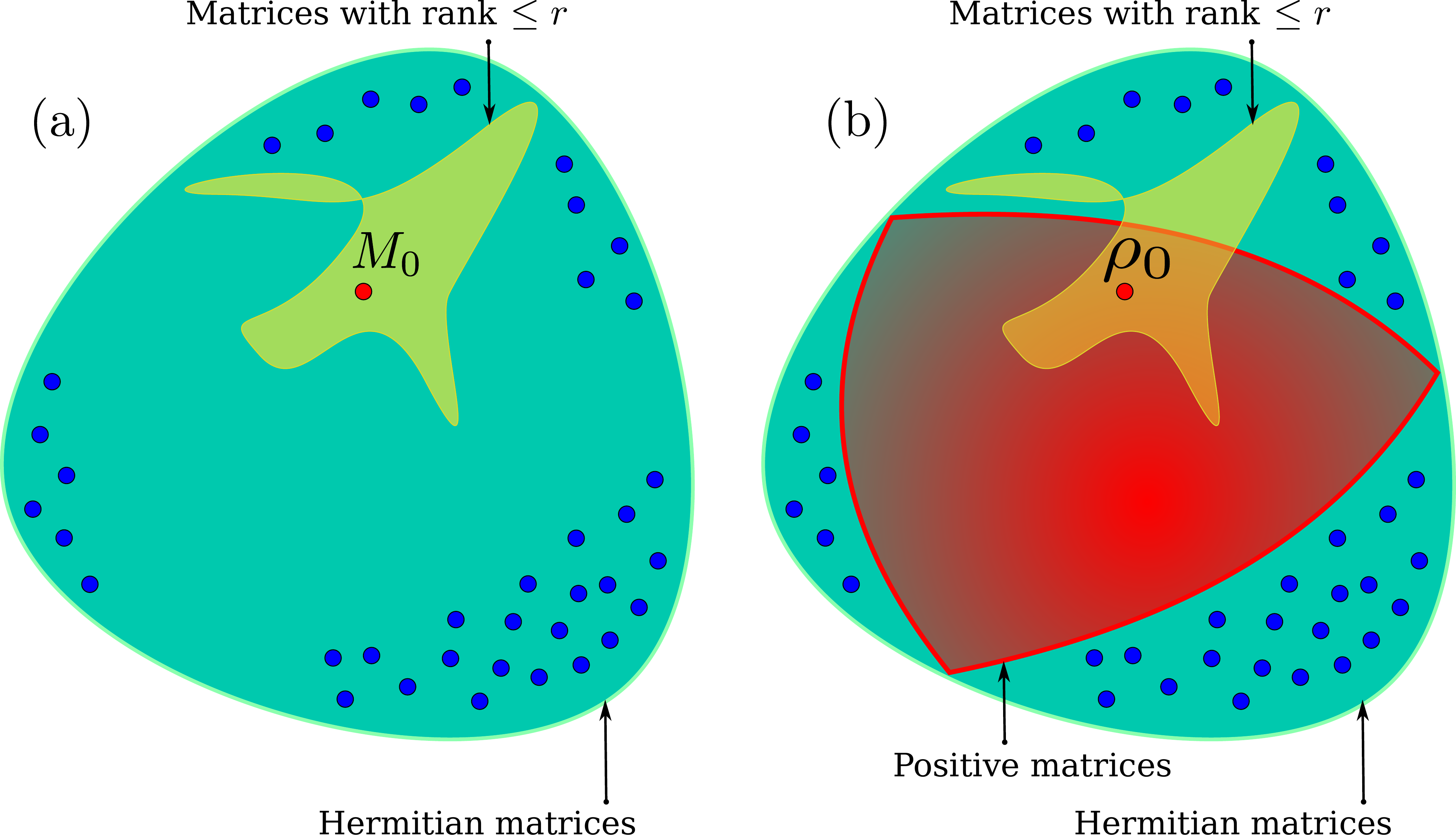}
\caption{{\bf Schematic illustration of Theorem~1.} (a) {\bf A generic compressed sensing scenario.} The noiseless measurement record uniquely specifies the low-rank signal matrix $M_0$ (represented by a red dot) within the set of Hermitian matrices with rank $\leq r$ (the yellow non-convex set). However, there are many other Hermitian matrices with rank larger than $r$ that are consistent with the measurement record (shown as the blue dots). (b)  {\bf A generic scenario of compressed sensing of quantum states.}  If the noiseless measurement record comes from a density matrix, i.e.,  a positive matrix $\rho_0\geq0$  whose rank $\leq r$, then, according to Theorem~1, it specifies $\rho_0$ uniquely among the set of  all positive matrices (shown as the red convex set). All other matrices that are consistent with the measurement record necessarily have negative eigenvalues and their rank is strictly larger than $r$.}
\label{fig:sets}
\end{figure}

Theorem~1 differs qualitatively from the standard compressed sensing theorem in a few key aspects. As discussed above, the general theory of compressed sensing guarantees that if the signal is a low-rank matrix with rank $\leq r$, and if the sensing map corresponds to compressed sensing measurements, then the measurement record uniquely specifies the unknown signal-matrix, but only within the subset of matrices with rank $\leq{r}$. Theorem~1, on the other hand,  states that if the matrix  to be estimated is constrained to be a positive matrix (e.g., a density matrix), then the measurement record uniquely specifies the matrix from within the  entire set of positive Hermitian matrices.  Therefore, without the positivity constraint, compressed sensing measurements correspond to rank-$r$ restricted-IC measurements of Eq.~\eqref{restricted_IC_definition}, whereas under positivity, the same measurements correspond to rank-$r$ strictly-IC measurements of Eq.~\eqref{strictly_IC_definition}. This central result of Theorem~1 is illustrated in Fig.~\ref{fig:sets}.

The implication of Theorem~1 for quantum-state tomography is as follows. Suppose that the state of the system $\rho_0$, a positive semidefinite matrix, has rank $\leq r$. Assume that we have measured the system with a sensing map that satisfies the appropriate compressed sensing property, and obtained the (noiseless) measurement record  ${\cal A}[{\rho_0}]=\bm{p}$. Then, according to Theorem~1,  ${\rho_0}$ is the only density matrix within the set of positive Hermitian matrices of any rank that yields the measurement probabilities $\bm{p}$.  Geometrically, as observed in~\cite{candes13,demanet14}, Theorem~1 states that the rank-deficient subset of the positive matrices cone is ``pointed.''  Therefore, under the promise that $\rk(\rho_0)\leq r$ and  ${\cal A}$ corresponds to compressed sensing measurements, the space of matrices $\rho$ that satisfy ${\cal A}[\rho]=\bm{p}$ and the cone of positive matrices intersect in a single point  $\rho=\rho_0$.

Theorem~1 implies that the solution set contains only one matrix, the density matrix $\rho_0$.  It follows that we can use any optimization method to search for it, and we are guaranteed to find it.  Thus, we have the following result:  Given a quantum measurement record  $\bm{p}={\cal A}[{\rho_0}]$, such that $\rk(\rho_0)\leq r$,  and where ${\cal A}$ corresponds to compressed sensing measurements, then the solution to 
\begin{equation}
\label{general_positive_CS}
\hat{\rho} = \arg\min_\rho \mathcal{C}(\rho)\;\; {\rm s.t.}\; {\cal A}[\rho]=\bm{p}\, \,  {\rm and} \, \,  \rho \geq 0,
\end{equation}
or to
\begin{equation}
\label{general_norm_positive_CS}
\hat{\rho} = \arg\min_\rho \Vert{\cal A}[\rho]-\bm{p}\Vert\;\; {\rm s.t.}\; \rho \geq 0,
\end{equation}
where $\mathcal{C}(\rho)$ is a any convex function of $\rho$, and $\Vert\cdot\Vert$ is any norm function, is unique:  $\hat{\rho} = \rho_0$. 
By confining the feasible set of matrices to positive matrices, we ensure that the measurement record uniquely identifies $\rho_0$ from the set of all density matrices, and thus any convex function of $\rho$ or the measurement error may serve as a cost function.   For example, this result applies to maximum-(log)likelihood estimation~\cite{hradil97} where $\mathcal{C}(\rho) =-\log( \prod_{\mu}\tr(E_\mu\rho)^{p_\mu})$. We thus conclude that when the feasible set of density matrices is constrained to be physical (i.e., have positive eigenvalues), any quantum tomography protocol whose sensing map corresponds to compressed sensing measurements will exhibit the compressed sensing effect.  We do not include a trace constraint in the convex programs above.  In the noiseless case considered here it is redundant.   Because the data came from a trace-preserving quantum measurements, the unique solution must be a normalized quantum state.  As discussed in the Supplementary information, the constraints $\rho\geq0$ and $\tr\rho=1$, taken together, immediately imply that $\rho_0$ is the only density matrix consistent with the noiseless data.  When we consider the important case of noisy measurements, the consequence trace constraint is nontrivial, as we discuss in the next section.\\

\begin{center}{\bf Robustness to measurement noise and model imperfection}\end{center}

So far, we have discussed the ideal case of a noiseless measurement record, where in the context of quantum tomography, $\bm p$ denoted a probability vector. The compressed sensing methodology, however, assures a robust reconstruction  of the signal in the presence of measurement noise. Our analysis inherits this crucial feature.  In a realistic scenario, we allow for a noisy measurement record, $\bm{f}={\cal A}[{\rho_0}]+\bm{e}$, where we assume that the noise contribution can be bounded by some norm $\Vert\bm{e}\Vert\leq\epsilon$.  In the context of quantum tomography we consider $\bm f$ to denote a vector of the observed frequencies of measurement outcomes. 

Theorem~1 ensures robust recovery of the positive density matrix if the noise level is small by solving any convex optimization problem.  Under the assumptions of Theorem~1, any convex minimization problem that searches for a solution within the cone of positive matrices must yield a solution $\hat\rho$ such that $\Vert\hat\rho-\rho_0\Vert\leq g(\epsilon)$, where $g(\epsilon)\rightarrow0$ as $\epsilon\rightarrow0$. From a geometrical point of view, when the noisy data arises from a rank-deficient state, as we gain data, there are fewer states that could have given rise to that data because the convex set of physical states is highly constrained near the point.  In the idealized limit of noiseless data, there is only one state compatible with the data.   Therefore, qualitatively, we expect a compressed sensing effect no matter how we search for the solution whenever the data arises from low rank positive matrices.   Quantitatively, of course, different heuristics may perform differently, yielding different estimates. Choosing the best optimization depends, in part, on the specific noise model. For example, in Supplementary information Section~B we derive a specific bound on the Frobenius (Hilbert-Schmidt) norm $\Vert\hat\rho-\rho_0\Vert_{\rm F}=\sqrt{\tr(\hat\rho-\rho_0)^2}$, where $\hat\rho$ is the solution of a nonnegative least-squares  program
\begin{equation}
\label{norm_positive_CS}
\hat{\rho} = \arg\min_\rho \Vert {\cal A}[\rho]-\bm{f}\Vert_2\;\; {\rm s.t.}\; \rho \geq 0.
\end{equation}
Whereas the normalization constraint, $\tr\rho = 1$ was unecessary in the noiseless case, in the case of a noisy measurement record, the convex optimization is not guaranteed to obtain a normalized state. One can include the trace constraint in the optimization, but it is generally unnecessary in the noisy case as well.  In fact, sometimes one can actually improve the robustness to noise by choosing $\tr\rho \neq 1$, as we discuss below.  In general, the output of the optimization should then be renormalized to give the final estimate.

We see this explicitly in~\cite{gross10}, where Gross {\em et al.} obtained a compressed sensing version of quantum-state tomography by solving the minimization problem,
\begin{equation}\label{QST NNTR}
\min_{\rho}\!{.} \;  \tr {\rho} \;\; {\rm s.t.}\; \Vert\bm{f} - {\cal A}[{\rho}]\Vert_2\leq\epsilon,\,{\rho}\geq0.
\end{equation}
This is equivalent to minimizing the nuclear norm of $\rho$ under the same constraints, i.e., when the feasible set is $\rho\geq 0$, then $\Vert\rho\Vert_* = \tr\rho$.  As noted above, minimizing the trace of the matrix in the absence of the positivity constraint is not equivalent to minimizing the nuclear norm, and therefore,  would not achieve compressed sensing.  While both Eq.~\eqref{norm_positive_CS} and  Eq.~\eqref{QST NNTR} are compressed sensing programs, in general they return different estimations.  However, in Supplementary information Section~C we show that the nonegative least-squares program,
\begin{equation}\label{cls}
{\min_{\rho}}.\;\Vert{\cal A}[{\rho}]-\bm{f}\Vert_2 \;\; \textrm{s.t.} \;\tr{\rho}= t,\, \rho \geq 0
\end{equation}
is exactly equivalent to the nuclear-norm minimization of Eq.~\eqref{QST NNTR} for a particular choice of $t$.  This fact was observed empirically in a recent experiment by Smith {\em et al.}~\cite{smith13}, in which quantum-state tomography via continuous measurement was achieved at a equivalent rate by both least-squares and trace minimization, with the positivity constraint included.  The difference between the final estimate was attributed to a difference in the robustness of the two estimators to noise.  Since Eqs.~\eqref{QST NNTR} and~\eqref{cls} are formally equivalent, the noisy measurement can be equivalently accommodated by solving~\eqref{cls} with a choice of $t$ that depends on the noise bound $\epsilon$.  As always, we renormalize to obtain the final density matrix.

In addition to noise in the measurements, there can be imperfections in the model.  When the sensing map satisfies the restricted isometry property, the compressed sensing methodology is not restricted to exact rank-deficient signal matrices.  It also ensures the robust recovery of the dominant rank-$r$ part of the density matrix. Our analysis shares this important and nontrivial property.  Lemma~2 given in Supplementary information Section~B is the root of this feature.

We have shown that Theorem~1 implies that for a positive matrix recovery, compressed sensing measurements correspond to a stronger notion of informationally completeness---a rank-$r$ strictly-IC. This implies that for quantum tomography we can construct  robust measurements that are also rank-$r$ strictly-IC. The robustness to measurement noise and model imperfection is guaranteed by the compressed sensing theory.  For example, in the context of a many-qubit system, Liu~\cite{liu11} showed that   ${\cal O}(rd{\rm~poly}(\log{d}))$ expectation values of Pauli products,  $w= \otimes_{i=1}^n \sigma_{\alpha_i}$, where $\sigma_{\alpha} \in \{I, \sigma_x, \sigma_y, \sigma_z\}$,  satisfy the restricted isometry property with overwhelming probability. Therefore, this set of expectation values is, with high probability,  a robust rank-$r$ strictly-IC measurement record.  Similar results hold for sparse quantum process matrix reconstruction, e.g., it is shown in~\cite{shabani11} that if the sensing map is constructed from random input states, and random observables, then the restricted isometry holds with high probability.

\begin{center}{\bf Numerical test: Compressed sensing state tomography of $n$-qubit system.}\end{center}

Gross and coworkers~\cite{gross10,liu11,flammia12} studied the problem of quantum-state tomography of an $n$-qubit system and showed that  $m={\cal O}(rd{\rm~poly}(\log{d}))$ expectation values of random Pauli observables satisfy an appropriate restricted isometry property with high probability. If these expectation values are obtained through measurements in Pauli bases, i.e., local projective measurements on individual qubits in the eigenbasis of the Pauli observables, then, in fact, we obtain much more information.  In addition to the average values, we also obtain the frequency of occurrence of each outcome,  $E_\mu  = \otimes_{i=1}^n  P_{\alpha_i}$, where $\mu$ indexes the series of $\alpha_i$, $\alpha=x,y,z$, and $ P_{\alpha_i} \in \{ \ket{{\up_{\alpha_i}}}\bra{{\up_{\alpha_i}}},  \ket{{\down_{\alpha_i}}}\bra{{\down_{\alpha_i}}}\}$.  Thus, we expect that we can obtain the required information for high-fidelity reconstruction using substantially fewer measurements based on individual outcomes in random Pauli bases rather than expectation values, and further reduce the resources needed for quantum-state tomography of a collection of qubits.

\begin{figure}
\centering
\includegraphics[width=0.8\linewidth]{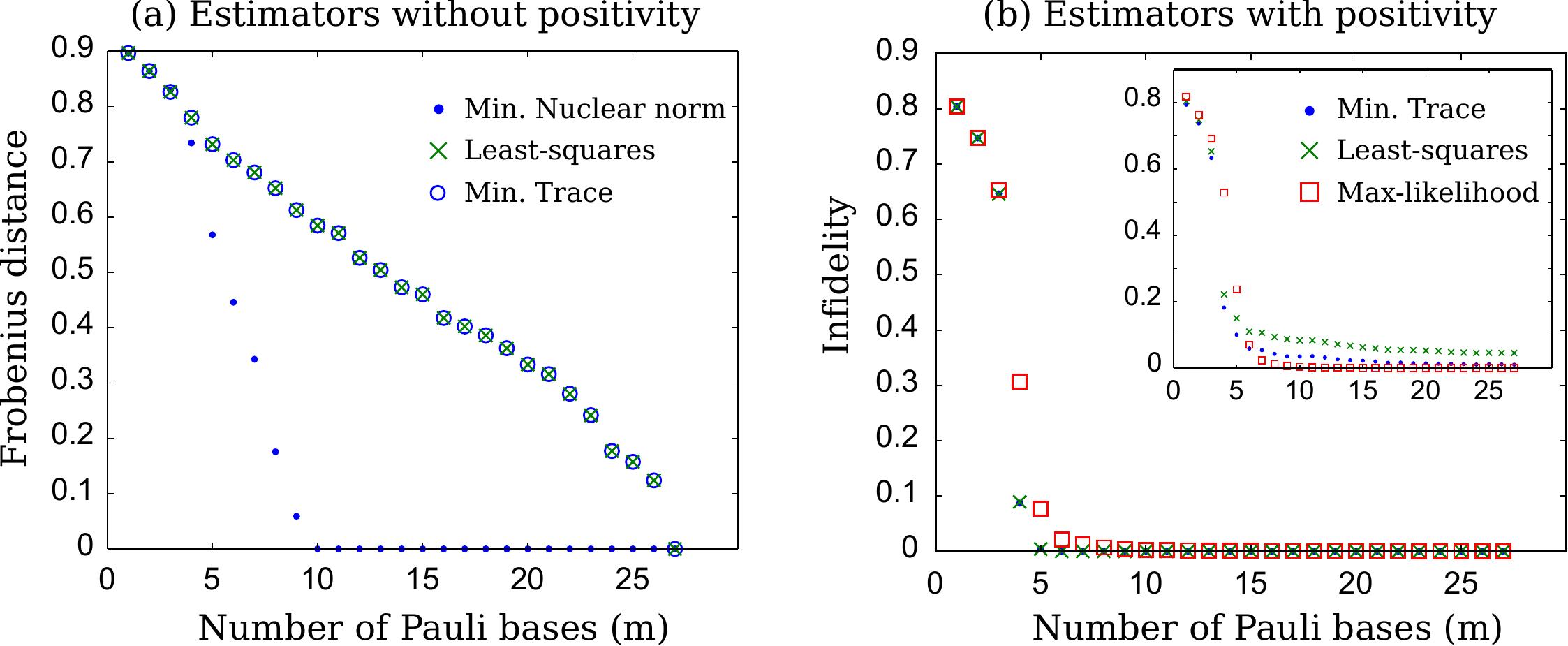}
\caption{{\bf Comparison of different estimators with and without the positivity constraint.} We simulate a three qubit system in which we produce a pure state, $\ket{\psi_0}$, and perform random projective measurements in the Pauli basis (all simulations are averaged over 10 Haar-random states). (a) {\bf Estimation without the positivity constraint.} We consider here the ideal case of a noiseless measurement record and plot the Frobenius distance between state $\ket{\psi_0}$ and the solution of an estimator, $\hat{\rho}$,  $\sqrt{\tr(\hat\rho-\ket{\psi_0}\!\bra{\psi_0})^2}$. The estimations are obtained by solving three different convex optimizations: (i) Nuclear-norm minimization: Eq.~\eqref{CS_matrix}, (ii) Least-squares minimization: $\hat{\rho}=\arg\min_{M} \Vert \bm{p}-{\cal A}[M]\Vert_2$, and  (iii) Trace minimization: $\hat{\rho}=\arg\min_M \tr(M) \,\, \, {\rm s.t.}\, \, \bm{p}={\cal A}[M]$.  The figure clearly shows that only nuclear-norm minimization achieves compressed sensing, i.e.,  exact recovery of the density matrix with a small number of measurement bases (here $m=10$).  Least-squares and  trace minimization require a full informationally complete measurement record with 27 Pauli bases to achieve exact recovery. (b) {\bf Estimation with the positivity constraint.} We plot here the  infidelity between $\ket{\psi_0}$ and the solution of different estimators,  $1 - \bra{\psi_0} \hat\rho\ket{\psi_0}$. The estimations are obtained by solving three different convex optimizations where the feasible set is constrained to the cone of positive matrices: (i) Nonnegative trace minimization (equivalently nuclear-norm minimization), Eq.~\eqref{QST NNTR}  (ii) Nonnegative least-squares minimization, Eq.~\eqref{norm_positive_CS}, and  (iii) The maximum-(log)likelihood estimator based on the algorithm described in~\cite{teo11}. In the main plot we simulate the case of an ideal noiseless measurement record;  in the inset plot we simulate a statistically noisy measurement record that corresponds to frequency of outcomes for $N_{\rm rep}=200$ repetitions. This figure exemplifies that when restricted to the set of positive matrices, all estimators are compressed sensing estimators.}
\label{fig:comparison}
\end{figure}

To exemplify this and the implication of Theorem~1, we perform numerical experiments on an $n$-qubit system (see Methods for details).  In Fig.~\ref{fig:comparison}, we simulate measurements on a three qubit system, $d=8$, and compare different numerical programs to estimate the state. In Fig.~\ref{fig:comparison}a  we solve three estimators:  Eq.~\eqref{CS_matrix} (nuclear-norm minimization),  $\min_M \Vert \bm{p}-{\cal A}[M]\Vert_2$ (least-squares), and $\min_M \tr(M) \,\, \, {\rm s.t.}\, \, \bm{p}={\cal A}[M]$ (trace minimization). Note that none of these estimators constrain the feasible set to the cone of positive matrices.  The least-squares and trace minimization are not convex heuristics for rank minimization, and thus, as expected, they do not achieve compressed sensing.  These programs require a full informationally complete measurement record in order to reconstruct the quantum state. On the other hand, the nuclear-norm heuristic does exhibit the compressed sensing  effect, and recovers the density matrix  with far fewer measurement outcomes.  In Fig.~\ref{fig:comparison}b we use the same data as in Fig.~\ref{fig:comparison}a, but here we use estimators that restrict the feasible set to positive semidefinite matrices, e.g.,  the nonnegative least-squares  estimator, Eq.~\eqref{norm_positive_CS}. The plots clearly show the implication of Theorem~1. Once restricted to the positive cone, the performance of all of the estimators is qualitatively the same and they all exhibit the compressed sensing effect. When the number of Pauli bases satisfy the appropriate restricted isometry property,  the various estimators find the exact state in the idealized situation where the measurement record has no noise, and they find a good estimate that is close to the true state of the system in the presence of noise  due to finite sampling statistics of $200$ repetitions.  

\begin{figure}
\centering
\includegraphics[width=0.5\linewidth]{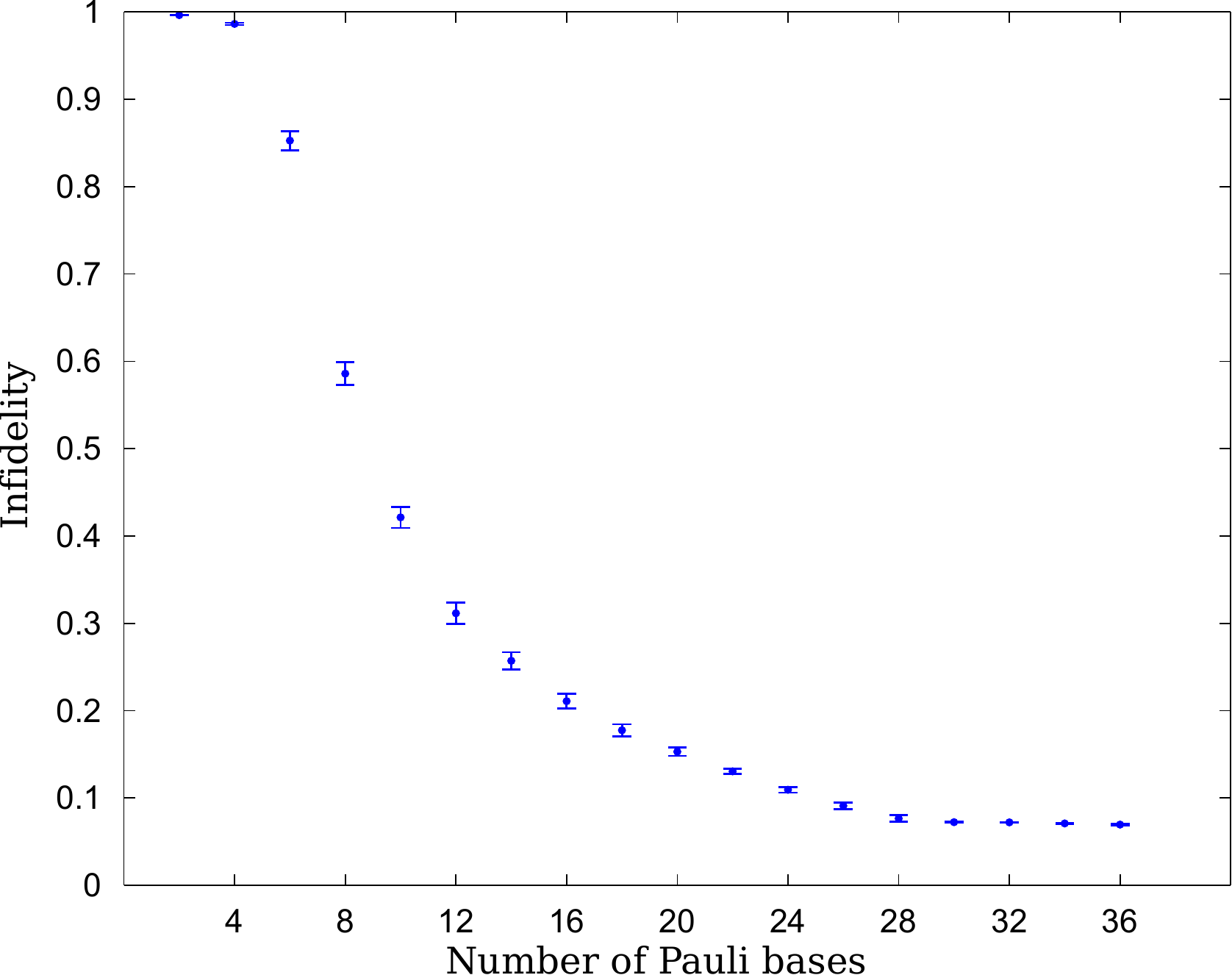}
\caption{{\bf Ten-qubit state tomography.} We simulate data based on random Pauli-projective measurements (see text).  The quantum tomography employs nonnegative least-squares, according to Eq.~\eqref{norm_positive_CS}. This algorithm can efficiently handle large dimensional matrices~\cite{carlos}.  We show the infidelity as a function of the number of measurement settings averaged over 10 Haar-random pure states (error bars shown).  The simulation clearly exhibits the compressed sensing effect. }
\label{fig:tenQubits}
\end{figure}

In Fig.~\ref{fig:tenQubits}, we treat a large dimensional Hilbert space: a ten qubit system, $d=2^{10}=1024$. We simulate 30 random Pauli bases of a Haar-random pure state with $N_{\rm rep}=100d$ repetitions for each observable.  We estimate the state by solving Eq.~\eqref{norm_positive_CS} with a convex optimization program that can efficiently handle such large dimensional data sets~\cite{carlos}.  The program implements a standard algorithm that uses gradient methods together with projection onto the positive cone. In the plot we see the compressed sensing effect due to the positivity constraint -- all the information is captured in about 28 random Pauli bases, given sufficient statistics.

\section{Discussion}

We have established a rigorous connection between the positivity property of quantum states and  the compressed sensing method for quantum tomography. Thanks to the positivity constraint associated with physical states, the record of such measurements allows for a unique identification of a low rank quantum state within the set of all physical quantum states, of any rank.  Thus, the measurements used for compressed sensing are informationally complete in a strict sense (strictly-IC). This aspect is fundamentally different than what happens if positivity is not included. In the absence of the positivity constraint, the compressed sensing measurements are informationally complete  in a restricted sense since they only uniquely identify a signal matrix from within the set of low rank matrices (restricted-IC). 

This strict relation has theoretical and practical implications. Most importantly, it implies that if one employs an optimization program that searches for a physical (positive) quantum state, any quantum tomography procedure whose sensing map corresponds to compressed sensing measurements will exhibit the compressed sensing effect.  This unifies apparently distinct numerical procedures such as maximum-likelihood and nuclear-norm minimization under the umbrella of compressed sensing.  From a practical perspective, when the positivity constraint is included, one can achieve compressed sensing estimation with any efficient convex optimization, such as ADMM algorithms~\cite{admm} developed to handle large dimensional matrices. 

Since the compressed sensing measurements  that satisfy the restricted isometry property are robust to measurement noise and model imperfection, this allows us to construct strictly-IC measurements that are robust against such noise. That is, if there is measurement noise and/or we strictly violate the assumption that rank$(\rho) \leq r$ (but only require that the density matrix is close to a density matrix with rank $\leq r$), then we are guaranteed that the estimation will be close to the unknown matrix.

Finally, though we have presented our results in the context of quantum-state tomography they are general and apply to the case of positive sparse vectors and positive rank-deficient matrices, the latter exemplified by quantum-process tomography.

\section{Methods}

\noindent{{\bf Compressed sensing measurements in matrix reconstruction.}}
A sensing map for matrix reconstruction, $\mathcal{A}$, is defined as a vector-valued linear map on a $d\times d$ Hermitian matrix, $y_i = {\cal A}[M]_i$.  This yields ``compressed sensing measurements for rank-$r$" if it guarantees  a robust recovery of matrices with rank $\leq r$ by solving a nuclear-norm minimization program, e.g., the compressed sensing heuristic, 
\begin{equation}\label{CS_heuristic}
\hat{M} = \arg\min_{M} \Vert{M}\Vert_*\,\, \, {\rm s.t.}\, \, \Vert{\cal A}[M]-\bm{f}\Vert_2\leq\epsilon,
\end{equation}
where $\bm{f}$ is the noisy measurement record, $\bm{f}=\bm{y}+\bm{e}$. When the matrix is promised to have rank $r$, the number of sufficient samples is of order $\mathcal{O}(rd)$, with possible logarithmic corrections, and the distance between the reconstruction $\hat{M}$ and $M_0$ is  $\mathcal{O}(\epsilon)$, where $\Vert{\bf e}\Vert_2\leq\epsilon$.  In this sense, the reconstruction is ``robust," and compressed sensing when $r\ll d$.  An analogous definition holds in the case of sensing maps for sparse vector reconstruction.

A sufficient condition that a sensing map yields compressed sensing measurements for matrix reconstruction is if it satisfies the ``restricted isometry property."  The map satisfies the restricted isometry property for rank-$r$ if there is some constant $0 \leq \delta_r <1$ such that, 
\begin{equation}\label{rip_matrix}
(1-\delta_r)\Vert {M}\Vert_{\rm F}^2\leq \Vert {\cal A}[{M}]\Vert_2^2\leq(1+\delta_r)\Vert {M}\Vert_{\rm F}^2,
\end{equation}
holds for all Hermitian matrices $M$ with rank $\leq r$, where $\Vert{M}\Vert_{\rm F} = \sqrt{\tr (M^\dag M)}$. The smallest constant $\delta_r$ for which this property holds is called the restricted isometry constant.

With small isometry constant $\delta_r$,  the sensing map ${\cal A}$ acts almost like an isometry when applied to rank $\leq r$ matrices, and thus allows us to effectively invert the measurement data to determine the matrix. Depending on the context, there are various results in the compressed sensing literature that apply for different values of the isometry constant. For example, Cand{\'e}s and collaborators~\cite{candes11}, show that the compressed sensing theory is applied when  $\delta_{4r}<\sqrt{2}-1$ (see Supplementary information Section~B).   

Our results are general and apply whenever the sensing map corresponds to compressed sensing measurements that assures robust recovery through the solution of Eq~\eqref{CS_heuristic}. While the  restricted isometry property is sufficient, our results are applicable in other cases, such as those described in~\cite{kueng14} where a robust recovery is guaranteed by ${\cal O} (r d)$ generic rank-one projections, or by ${\cal O} (r d \log(d))$ projectors onto random elements of an approximate 4-design. \vspace{0.3cm}

\noindent{{\bf Numerical experiments.}}
In our numerical experiments, we simulate independent measurements of random Pauli bases on a Haar-random pure state of dimension $d=2^n$, $\rho_0=\ket{\psi_0}\bra{\psi_0}$. The measurement record, given by the frequency of outcomes, $\bm{f}$, is generated by sampling $N_{\rm rep}$ times from the probability distribution $\bm{p}=\tr(\bm{E}\rho_0)$.  Here $\bm{E}$ is the vector of POVM elements, each corresponding to a tensor product of projectors onto the eigenbasis of Pauli observables, $E_\mu  = \otimes_{i=1}^n  P_{\alpha_i}$, where $\mu$ indexes the series of $\alpha_i$, $\alpha=x,y,z$, and $ P_{\alpha_i} \in \{ \ket{{\up_{\alpha_i}}}\bra{{\up_{\alpha_i}}},  \ket{{\down_{\alpha_i}}}\bra{{\down_{\alpha_i}}}\}$.   The measurement record is then used in various estimators~\cite{cvx}. We measure the performance by the average infidelity over 10 random pure states, $1-\overline{\bra{\psi_0} \hat{\rho} \ket{\psi_0}}$. 

\vspace{0.5cm}

\noindent{{\bf Acknowledgments}}\\
We thank Jens Eisert and  Carlos A. Riofr\'{\i}o for stimulating discussions. In particular we thank J.E. for his insights regarding the proof of Theorem~1 and C.A.R. for initial work that led to Supplementary information Section~C and the development of numerical methods used here. A.K. and I.H.D. acknowledge the support of NSF Grants PHY-1307520 and PHY-1212445.  R.L.K partially supported by the ARO MURI grant W911NF-11-1-0268.\\

\noindent{{\bf Author contributions}}\\
All authors contributed ideas. A.K. performed the calculations.  All authors wrote the manuscript.\\

\noindent{{\bf Additional information}}\\
Correspondence and
requests for materials should be addressed to A.K.\\

\noindent{{\bf Competing financial interests}}\\
The authors declare no competing financial interests.

\appendix
\section*{Supplementary Information}
\begin{center}{\bf Section A: Proof of Theorem~1}\end{center}
In a direct extension of Bruckstein {\em et al.}~\cite{bruckstein08}, we first show that under the appropriate conditions, positivity implies that the set $\{M \vert {\cal A}[{M}]=\bm{p},\; M\geq0\}$ contains a single element. The proof does not assumes that $M$ is a quantum state nor that the sensing map is related to a POVM of any kind. It applies for the general circumstances of positive matrices and sensing maps.

Consider the sensing map ${\cal A}[\odot]=\tr(\bm{E}\odot)$, where the elements of the vector $\bm{E}$ are some matrices, $E_\mu$, $\mu=1,2,\ldots,m$. Suppose that the span of $\{E_\mu \}$ is strictly positive, namely,   
\begin{equation*}
\exists\,\bm{h}=(h_1,h_2,\ldots,h_m)^\intercal\; {\rm s.t.}\; \bm{h}^\intercal\bm{E}\equiv\sum_{\mu=1}^m h_{\mu}E_\mu=W>0, 
\end{equation*}
with $W=BB^\intercal$ a $d\times d$ (strictly) positive matrix.   This allows us perform a change of representation to an auxiliary problem.  Defining, ${\cal D}[\odot]=\tr(B^{-1}\bm{{E}}B^{\intercal-1} \odot)$, and $Z{=}B^\intercal{M}B$, there is one-to-one correspondence between the solution sets 
\begin{equation*}
\{M \;\vert\; {\cal A}[M]=\bm{p},\;M\geq0\}\;{\rm and}\;\{Z\;\vert\; {\cal D}[Z]=\bm{p},\;Z\geq0\}, 
\end{equation*}
and the rank of the solutions are the same. An important property of the modified problem is that
\begin{equation*}
\tr (Z)=\tr (W{M})\comment{=\tr (\bm{h}^\intercal\bm{{E}}M)=\bm{h}^\intercal \tr (\bm{{E}}{M})}=\bm{h}^\intercal\bm{p}=c.
\end{equation*}
That is, the trace of $Z$ is fixed, and its value depends on $\bm{p}$ and the choice of $\bm{h}$. Therefore, we can refine the above statement:  there is a one-to-one correspondence between the solution sets 
\begin{equation*}
\{M \vert {\cal A}[M]=\bm{p},M\geq0\}\,{\rm and}\,\{Z\vert {\cal D}[Z]=\bm{p},Z\geq0,\tr Z=c\}, 
\end{equation*}
and the rank of the solutions are the same.\\

{\bf Lemma~1.} Assume $\bm{p}={\cal D}[Z_0]$ for some $Z_0\geq0$ with $\rk(Z_0)\leq r$. If ${\cal D}$ satisfies the restricted isometry property with constant $\delta_{4r}<\sqrt{2}-1$, then the set $\{Z\;\vert\; {\cal D}[Z]=\bm{p},\;Z\geq0,\;\tr Z=c\}$ contains only one element, $Z=Z_0$.
\begin{proof} 
The Lemma assumes that $\rk({Z_0})\leq r$ and  $\delta_{4r}<\sqrt{2}-1$.  Therefore, according to the Theorem 2.4 of~\cite{candes11} (applied to the noiseless case), the problem 
\begin{equation*}
\hat Z=\arg\min_{Z}\!{.} \;  \Vert{Z}\Vert_{*}\;\; {\rm s.t.}\;{\cal D}[{Z}]=\bm{p},
\end{equation*}
has a unique minimizer  $\hat Z=Z_0$. But since the feasible set contains only positive matrices, then $\Vert{Z}\Vert_{*}=\tr Z$. Therefore any other positive solution to ${\cal D}[{Z}]=\bm{p}$ must have a trace larger than  $\tr(\hat Z)=c$, thus it is necessarily not in the set $\{Z\;\vert\; {\cal D}[Z]=\bm{p},\;Z\geq0,\;\tr Z=c\}$. Hence, this set contains only one element, as claimed. 
\end{proof} 
Since, the set $\{Z\vert{\cal D}[Z]=\bm{p},Z\geq0,\tr Z=c\}$ contains only one element, so does the set $\{M \vert {\cal A}[M]=\bm{p},M\geq0\}$ given that ${\cal D}$ satisfies the restricted isometry property with constant $\delta_{4r}<\sqrt{2}-1$. In general, it is required to find  a transformation of the sensing map ${\cal A}$ that yields ${\cal D}$ with $\delta_{4r}<\sqrt{2}-1$.  

This general result can be applied to the specific case of quantum tomography, where now $M=\rho$, a positive-semidefinite density matrix, and the elements of the vector $\bm{E}$ form a (trace preserving) POVM. In this case, we can choose $\bm{h}=(1,1,\ldots,1)^{\intercal}$, a vector whose elements are all 1, then $W=\bm{h}^\intercal\bm{E}=\sum_\mu{E}_\mu=\mathbb{1}$, and thus ${\cal D}={\cal A}$. Therefore, in this particular case, $\delta_{4r}({\cal D})=\delta_{4r}({\cal A})$. 

Note that in order to show the generality of our result in CS, we have chosen to present arguments in the course of the proof that apply to general positive matrices and sensing maps and only then to apply it specifically to the quantum tomography case. In the quantum case, however, this theorem follows directly, without the need for the construction of Bruckstein {\em et al.}.  For a trace-preserving POVM, it follows immediately that ${\cal D}={\cal A}$ and $Z=\rho$. Therefore, for quantum tomography all of the arguments above that are made with relation to ${\cal D}$ and $Z$ can be made on ${\cal A}$ and $\rho$ directly, and Theorem 1 follows as extension of~\cite{candes11}, applied to positive matrices~\cite{jens}.\vspace{0.5cm}

\begin{center}{\bf Section B: Bound on $\Vert {\hat\rho}-{\rho_0}\Vert_F$}\end{center}
Consider the following heuristic
\begin{equation}\label{NN}
\hat{M}=\arg{\min_{M}}.\;\Vert {M}\Vert{_*} \;\; \textrm{s.t.} \;\Vert{\cal A}({M})-{\bm f}\Vert_2\leq\epsilon 
\end{equation}
Suppose that $M_0$ is an arbitrary rank matrix. Let $M_0=U{\rm diag}(\bm\sigma)V^*$ be the singular value decomposition of $M_0$ where $\bm\sigma$ is the list of ordered singular values $\sigma_1\geq\sigma_2\geq\cdots\geq\sigma_d$. We let $M_{r}$ be  the part of $M_0$ corresponding to its largest $r$ singular values. By definition $M_{\rm c}=M_0-M_r$ corresponds to the $d-r$ smallest singular values of $M_0$, i.e., the `tail' of $M_0$. 

To bound $\Vert {\hat\rho}-{\rho_0}\Vert_F$ we use the following Lemma.\\
{\bf Lemma~2.} Suppose $\delta_{4r} < \sqrt2-1$ and let $M_0$ be a matrix such that $\Vert{\cal A}(M_0)-\bm{f}\Vert_2\leq\epsilon$. Then the solution $\hat{M}$ to~\eqref{NN} obeys
\begin{equation}\label{lemeq:nn}
\Vert \hat{M}-M_0\Vert_{\rm F}\leq C_{0} \epsilon +C_1\sqrt{\frac2{r}} \Vert M_{\rm c}\Vert_{*}
\end{equation}
where $C_0=\frac{4\sqrt{1+\delta_{4r}}}{1-(1+\sqrt{2})\delta_{4r}}$, and ${C}_1=\frac{1-(1-\sqrt{2})\delta_{4r}}{1-(1+\sqrt{2})\delta_{4r}}$ are constants depending only on the isometry constant $\delta_{4r}$.\\

Lemma~2, is somewhat different than Lemma~3.2 proved in~\cite{candes11}. However the proof of Lemma~3.2 applies directly to  Lemma~2

An important special case of this  Lemma~2 is for a signal matrix $M_0$ with $\rk(M_0)\leq{r}$, that satisfies $\Vert{\cal A}(M_0)-\bm{f}\Vert_2\leq\epsilon$. For this case, $M_{\rm c}=0$, and therefore
\begin{equation}\label{thm:nn}
\Vert \hat{M}-M_0\Vert_{\rm F} \leq C_0\epsilon.
\end{equation}

We are now ready to bound $\Vert {\hat\rho}-{\rho_0}\Vert_F$. Using the triangle inequality and the result of equation~\eqref{thm:nn} we get
\begin{align}\label{MnnlsMtri}
\Vert {\hat\rho}-{\rho_0}\Vert_F &\leq \Vert {\hat\rho}-{\rho}^*\Vert_F+\Vert {\rho}^*-{\rho_0}\Vert_F\nn
&\leq \Vert {\hat\rho}-{\rho}^*\Vert_F+C_0\epsilon.
\end{align}
where ${\rho}^*$ is the solution for equation~\eqref{NN}, and $C_0=\frac{4\sqrt{1+\delta_{4r}}}{1-(1+\sqrt{2})\delta_{4r}}$.

To bound $\Vert {\hat\rho}-{\rho}^*\Vert_F$ we use the result of  Lemma~2 which give an upper bound on $\Vert {\rho}-{\rho}^*\Vert_F$. The only assumption regarding ${\rho}$ that entered the proof of Lemma~2 is that it is a feasible matrix, $\Vert {\cal A}({\rho})-\bm{f}\Vert\leq\epsilon$.  However, ${\hat\rho}$ is a feasible matrix for the problem of~\eqref{NN} since by its definition  it minimizes $\Vert {\cal A}(\cdot)-\bm{f}\Vert$. Therefore, necessarily,  $\Vert {\cal A}({\hat\rho})-\bm{f}\Vert\leq\epsilon$.

Applying the result of Lemma~2 to bound $\Vert \hat\rho-{\rho}^*\Vert_F$, we can rewrite inequality~\eqref{MnnlsMtri} as 
\begin{align*}
\Vert {\hat\rho}-{\rho_0}\Vert_F &\leq 2C_0\epsilon+C_1\sqrt{\frac2{r}}\Vert ({\hat\rho})_{\rm c}\Vert_{*}\nn
\end{align*}
where ${C}_1=\frac{1-(1-\sqrt{2})\delta_{4r}}{1-(1+\sqrt{2})\delta_{4r}}$.\vspace{0.4cm}

\begin{center}{\bf Section C: Proof of formal equivalence between equation~(8) and equation~(9) of the main text}\end{center}
Consider the two minimization programs
\begin{align}\label{convexTR}
p_{\rm tr}=\min_{\rho}\!{.} &\;  \tr {\rho} \nn
\textrm{subject to:} &\;\Vert\bm{f} - {\cal A}({\rho})\Vert_2\leq\epsilon 
\end{align}
and
\begin{align}\label{convexLS}
p_{\rm ls}=\min_{\rho}\!{.} &\; \Vert\bm{f} - {\cal A}({\rho})\Vert_2\nonumber \\
\textrm{subject to:} &\;\; \tr{\rho}= t,
\end{align}
where as before, ${\cal A}$ is a linear map, ${\cal A}:{\mathbb R}^{d\times d}\rightarrow{\mathbb R}^{m}$ and $\bm{f}$ is the record $\bm{f} ={\cal A}({\rho_0})+\bm{\e}$, where ${\rho}$ is the density matrix and $\bm{\e}$ denotes the noise. Similarly to Ref.~\cite{candes11}, we take the map to be of the form  ${\cal A}({\rho})=\tr(\bm{E}{\rho})$, where $\bm{E}=(E_1,E_2,\ldots,E_m)$, and $E_j$, $j=1,2,\ldots,m$, are $d\times d$ matrices represent the measurement operators. Inspired by the formulation of measurement we further assume that $E_j\geq0$ and $\sum_j E_j=1$.\\
{\bf Lemma~3} For a given map ${\cal A}$ and a record $\bm{f}$,  if $t= \sum_{j=1}^{m} f_{j}-\sqrt{m}\epsilon$, then the two convex programs~\eqref{convexTR} and~\eqref{convexLS} are the mathematically equivalent.
\begin{proof}
Since the objective functions and the constrains of the two convex programs are linear or quadratic, both programs have zero duality gap, thus a strong duality holds for them both. To prove the Lemma we construct and solve the dual problem of each (primal) program and than show that for $t=\sum_{j=1}^{m} f_{j}-\sqrt{m}\epsilon$ the solutions of the two corresponding dual problems coincide. Since there is no duality gap for these problems, this implies that the  solutions to the two primal problems, first, equal to the solutions of the dual problems and, second, coincide with each other, as claimed. 

The (conic) Lagrangian of~\eqref{convexTR} is given by,
\begin{equation}
L({\rho},\bm{u},\lambda)=\tr{\rho} +\sum_{j} u_j (f_{j} - \textrm{Tr} (E_{j} {\rho})) -\lambda\epsilon,
\end{equation}
with the dual variable (Lagrange multipliers) $\Vert \bm{u}\Vert_2 \leq \lambda$, and $\lambda\geq 0$.
The dual function is obtained by $\min{\!.}_{\rho} L$, which is given by the condition $\nabla_{\rho} L=0$. Using $\nabla_{\rho} \tr {\rho}A=A$ we get
\begin{equation}\label{trCondition}
\nabla_{\rho} L=-\sum_{j} u_j {E}_{j} +\mathbb{1}=0\Rightarrow \sum_{j=1}^{{m}} u_j {E}_{j}=\mathbb{1},
\end{equation}
and therefore, 
\begin{equation}
\min_{\rho}\!{.}\; L=\sum_{j} u_j f_{j} -\lambda\epsilon.
\end{equation}
with $\Vert \bm{u}\Vert_2 \leq \lambda$, and $\lambda\geq 0$. The dual problem of~\eqref{convexTR} thus reads
\begin{align}\label{trDual}
d_{\rm tr}=\max_{\bm{u},\lambda}\!{.}&\; \sum_{j} u_j f_{j} -\lambda\epsilon\nn
\textrm{subject to:} &\;\; \Vert \bm{u}\Vert_2 \leq \lambda\nn 
&\; \lambda\geq 0
\end{align}
In fact we can solve this program exactly. equation~\eqref{trCondition}, $\sum_{j=1}^{{m}} u_j {E}_{j}=\mathbb{1}$, together with $\sum_{j=1}^{{m}} {E}_{j}=\mathbb{1}$ implies a solution $u_j =1$ for $j=1,\ldots,{m}$. Therefore the condition $\Vert \bm{u}\Vert_2 \leq \lambda$ now reads $\sqrt{{m}} \leq \lambda$. Moreover $\sum_{j} u_j f_{j}=\sum_{j} f_{j}$. Plugging all that in equation~\eqref{trDual}, we obtain
\begin{align}
d_{\rm tr}=\max_{\lambda}\!{.}&\; \sum_{j=1}^{m} f_{j} -\lambda\epsilon\nn
\textrm{subject to:} &\;\; \sqrt{{m}} \leq \lambda.
\end{align}
The solution of this problem is given by taking the minimum value of $\lambda$, $\lambda=\sqrt{{m}}$, that is, $d_{\rm tr}= \sum_{j=1}^{m} f_{j}-\sqrt{{m}}\epsilon$. Since we have a strong duality in this program we get that 
\begin{equation}\label{trDualSolution}
p_{\rm tr}=d_{\rm tr}=\sum_{j=1}^{m} f_{j}-\sqrt{{m}}\epsilon.
\end{equation}
Let ${\rho}_{\rm tr}$ be the argument that solves~\eqref{convexTR}, then,  
\begin{align}\label{rho_tr}
\tr {\rho}_{\rm tr}&={p}_{\rm tr}= \sum_{j=1}^{m} f_{j}-\sqrt{{m}}\epsilon.
\end{align}

Next, we consider the problem of~\eqref{convexLS} which is equivalent to,
\begin{align}
\min_{\rho}\!{.} \max_{\bm{v}}\!{.}&\; \langle \bm{v},\bm{f} - \textrm{Tr} (\bm{E} {\rho})\rangle\nonumber \\
\textrm{subject to:} &\;\; \tr{\rho}= t\nonumber \\
			  &\; \Vert \bm{v}\Vert_2 \leq 1. 
\end{align}
Thus, the (conic) Lagrangian function of this problem is  given by,
\begin{equation}
L({\rho},\bm{v},\mu)=\sum_{j} v_j (f_{j} - \textrm{Tr} ({E}_{j} {\rho}) )+\mu (\tr{\rho}-t),
\end{equation}
with  $\Vert \bm{v}\Vert_2 \leq 1$.
The dual function is obtained by $\min\!{.}_{\rho}L$, which is given by the condition $\nabla_{\rho} L=0$. Using $\nabla_{\rho} \tr{X} A=A$, we get
\begin{equation}\label{lsCondition}
\nabla_{\rho} L=-\sum_{j} v_j {E}_{j} +\mu\mathbb{1}=0\Rightarrow \sum_{j=1}^{{m}} v_j {E}_{j}=\mu\mathbb{1},
\end{equation}
and therefore, 
\begin{equation}
\min_{\rho}\!{.} L=\sum_{j} v_j f_{j} -\mu t
\end{equation}
with  $\Vert \bm{v}\Vert_2 \leq 1$. The dual problem thus reads
\begin{align}\label{lsDual}
d_{\rm ls}=\max_{\bm{v},\mu}\!{.}&\; \sum_{j} v_j f_{j} -\mu t\nn
\textrm{subject to:} &\;\; \Vert \bm{v}\Vert_2 \leq 1 
\end{align}
Similarly to the previous case, we can solve this program exactly. equation~\eqref{lsCondition}, $\sum_{j=1}^{{m}} v_j {E}_{j}=\mu\mathbb{1}$, together with $\sum_{j=1}^{{m}} {E}_{j}=\mathbb{1}$ implies a solution $v_j =\mu$ for $j=1,\ldots,{m}$. Therefore, the condition $\Vert \bm{v}\Vert_2 \leq 1$ now reads $\sqrt{{m}}\mu \leq 1$, that is, $\mu \leq 1/\sqrt{{m}}$. Moreover $\sum_{j} v_j f_{j}=\mu\sum_{j} f_{j}$. Plugging all that in equation~\eqref{lsDual}, we obtain
\begin{align}
d_{\rm ls}=\max_{\mu}\!{.}&\; \mu\Bigl(\sum_{j=1}^{m} f_{j}-t\Bigr)\nn
\textrm{subject to:} &\;\; \mu \leq 1/\sqrt{{m}}.
\end{align}
The solution to this problem is given by taking the maximum value of $\mu$, $\mu=1/\sqrt{{m}}$, i.e.,  $d_{\rm ls}=(\sum_{j=1}^{m} f_{j}-t)/\sqrt{{m}}$. Since we have a strong duality in this program we get that 
\begin{equation}\label{lsDualSolution}
p_{\rm ls}=d_{\rm ls}=\frac1{\sqrt{m}}\Bigl(\sum_{j=1}^{m} f_{j}-t\Bigr).
\end{equation}
Let ${\rho}_{\rm ls}$ be the argument that solves~\eqref{convexTR}, then $\tr {\rho}_{\rm ls}=t$ and $\Vert\bm{f} - \tr (\bm{E} {\rho}_{\rm ls})\Vert_2=p_{\rm ls}$. 

The mathematical equivalence between the two programs,~\eqref{convexTR} and~\eqref{convexLS}, is obtained  for taking $t$ in equation~\eqref{convexLS}  to be equal to $p_{\rm tr}=\tr {\rho}_{\rm tr}$. For this value of $t$, $t=\tr {\rho}_{\rm tr}$, we obtain 
\begin{align}\label{rho_ls}
d_{\rm ls}=&\Vert\bm{f} - \tr (\bm{E} {\rho}_{\rm ls})\Vert_2=\epsilon.
\end{align}

The problem of~\eqref{convexTR}  finds a matrix ${\rho}_{\rm tr}$ which has the minimal trace and satisfies $\Vert\bm{f} - \textrm{Tr} (\bm{E} {\rho}_{\rm tr})\Vert_2\leq\epsilon$. Using the value of $t=\tr {\rho}_{\rm tr}$ in~\eqref{convexLS}, means that the program~\eqref{convexLS} finds the matrix ${\rho}_{\rm ls}$ which has the minimal $\Vert\bm{f} - \textrm{Tr} (\bm{E} {\rho}_{\rm ls})\Vert_2$ and satisfies $\tr {\rho}_{\rm ls}=\tr {\rho}_{\rm tr}$. We showed that the solution is such that the minimal value is $\Vert\bm{f} - \textrm{Tr} (\bm{E} {\rho}_{\rm ls})\Vert_2=\epsilon$. This implies that every element in the set  $\{\rho\vert\tr {\rho}=\tr {\rho}_{\rm tr}\}$ satisfies $\Vert\bm{f} - \textrm{Tr} (\bm{E} {\rho})\Vert_2\geq\epsilon$. Therefore, we conclude that,  the solution of~\eqref{convexTR} necessarily satisfies $\Vert\bm{f} - \textrm{Tr} (\bm{E} {\rho}_{\rm tr})\Vert_2=\epsilon$.
This in turn imply that both programs~\eqref{convexTR} and~\eqref{convexLS} return the same solution $\hat{\rho}$ with $\tr{\hat\rho}= \sum_{j=1}^{m} f_{j}-\sqrt{{m}}\epsilon$ and $\Vert\bm{f} - \tr(\bm{E} {\hat\rho})\Vert_2=\epsilon$.
\end{proof}

The programs~\eqref{convexTR} and~\eqref{convexLS} with $t=\tr(\rho_{\rm tr})$ remain equivalent, upon adding any convex constraint to them both. In particular, the two programs 
\begin{align}\label{convexTRpos}
\min_{\rho}\!{.} &\;  \tr{\rho} \nn
\textrm{subject to:} &\;\Vert\bm{f} - {\cal A}({\rho})\Vert_2\leq\epsilon\nn
&\;\; {\rho}\geq0
\end{align}
and
\begin{align}\label{convexLSpos}
\min_{\rho}\!{.} &\; \Vert\bm{f} - {\cal A}({\rho})\Vert_2\nn
\textrm{subject to:} &\;\; \tr{\rho}= \tr(\rho_{\rm tr})\nn
&\;\; {\rho}\geq0
\end{align}
are mathematically equivalent as claimed. 

Lastly, we remark that the while the proof of equivalence was given here using the two-norm, $\Vert\cdot\Vert_2$, it holds for any norm.  Therefore, the mathematical equivalence between the programs of~\eqref{convexTRpos}, \eqref{convexLSpos} also holds if we replace the two-norm that appears in these programs by any other norm. 

\end{document}